\documentclass[10pt,twocolumn,letterpaper]{article}

\usepackage{iccv}
\usepackage{times}
\usepackage{epsfig}
\usepackage{graphicx}
\usepackage{amsmath}
\usepackage{amssymb}
\usepackage{multirow}
\usepackage{color,graphicx}
\usepackage{subfig}
\usepackage{booktabs}
\usepackage[table,xcdraw]{xcolor}
\usepackage{cite}
\usepackage{filecontents}
\usepackage{caption}
  
 \iccvfinalcopy 

\def\iccvPaperID{****} 

\ificcvfinal\fi

\begin{document}

\title{Random Hash Code Generation for Cancelable Fingerprint Templates using Vector Permutation and Shift-order Process}

\author{Sani M. Abdullahi and Shuifa Sun\\
College of Computer and Information Technology\\
China Three Gorges University, Yichang 443002, China\\
{\tt\small sani@my.swjtu.edu.cn, watersun@ctgu.edu.cn}
}

\maketitle
\ificcvfinal\thispagestyle{plain}\fi

\begin{abstract}
   Cancelable biometric techniques have been used to prevent the compromise of biometric data by generating and using their corresponding cancelable templates for user authentication. However, the non-invertible distance preserving transformation methods employed in various schemes are often vulnerable to information leakage since matching is performed in the transformed domain. In this paper, we proposed a non-invertible distance preserving scheme based on vector permutation and shift-order process. First, the dimension of feature vectors is reduced using kernelized principle component analysis (KPCA) prior to randomly permuting the extracted vector features. A shift-order process is then applied to the generated features in order to achieve non-invertibility and combat similarity-based attacks. The generated hash codes are resilient to different security and privacy attacks whilst fulfilling the major revocability and unlinkability requirements. Experimental evaluation conducted on 6 datasets of FVC2002 and FVC2004 reveals a high-performance accuracy of the proposed scheme better than other existing state-of-the-art schemes.
\end{abstract}

\section{Introduction}

Fingerprint authentication systems have witnessed a progressive deployment in various areas of our daily endeavors, stretching from a simple task of verifying ourselves on our smartphones to a broader area of identity management such as visa application and cross boarder checking procedures. Such huge reliant on fingerprints amongst other modalities has to do with its permanence, uniqueness, proven stability, assurance, individuality and cost-effectiveness\cite{jain2008}\cite{nalini2001}\cite{nandakumar2015}\cite{jain2016}. However, the stored fingerprint templates are vulnerable to attacks, and once compromised, there would be a permanent loss of vital data since it is irrevocable and irreplaceable. As such, the devastating consequence that comes with the loss is immeasurable.
To overcome these drawbacks, cancelable templates are used to efficiently preserve the security and privacy concerns of fingerprint biometric data by permitting compromised templates to be replaced by a new one. They are built on the idea that they can be revoked and a new instance of the corresponding templates can be re-generated if the original data is tampered with. This is often achieved through non-invertible transformation, i.e., an intentional (but crafted) distortion of the biometric data by means of transformation function and user-predefined parameters. The transformation function can be applied either in a form of blending mechanism (salting) or a many-to-one transformation (non-invertibility). The essential properties of cancelable biometric schemes are as follows:\\a)	Diversity: Different cancelable templates should be generated from same biometric data. In other words, same biometric data should not correlate in different applications.\\b)	Non-invertibility: It should be impossible to reconstruct the original biometric data from its corresponding transformed version even if the transformed templates and the transformation methods are known.\\c)	Revocability: It should be possible to revoke a compromised template and replace it with a new one based on the original data.\\d)	Performance: The accuracy performance should not be degraded after the template transformation. In other words, the incorporation of security mechanism should not reduce the systems’ recognition accuracy.

Generally, the accuracy performance of a cancelable biometric system is known to be directly impacted by the feature extraction and representation of the biometric data. Minutiae descriptors provides us with the efficient means of representing their discriminative features for effective construction of fingerprint templates. An example of such descriptor is the state-of-the-art minutiae cylinder code (MCC)\cite{RN24} which is known for its superior matching performance. However, this descriptor exhibits some security and privacy shortcomings as they are prone to inversion. On the other hand, the incorporation of security mechanism such as in the protected version of MCC, known as 2PMCC\cite{RN27}. The method is able to avert the inversion but still prone to revocability. Besides non-invertibility and revocability, the accuracy performance is often degraded after transformation due to modification of the minutiae descriptor, coupled with information leakage in the matching procedure which raise concerns over the systems’ full non-invertibility.

\section{Review works}

In the section, several works that focus on the non-invertibility of fingerprint templates including the recent distance preserving drawbacks in most of these approaches are discussed. Readers can as well refer to these papers \cite{RN37}\cite{RN36}\cite{nandakumar2015} for a more in-depth survey on cancelable biometrics.
The earlier works of \cite{RN27}\cite{RN35}\cite{RN38}, and \cite{RN32} are all designed with notion of achieving non-invertibility. However, even though it is not fully achieved due to the fact that the distance between features is not attained, their accuracy is also degraded considering that the security mechanisms are directly applied on the descriptors or incorporated in the mapping approach, except for \cite{RN27}. Even so, it also suffers accuracy degradation depending on the parameter tuning. Wang et al \cite{RN42} designed a cancelable fingerprint templates using zoned minutiae pairs. Local minutiae structures are leveraged to construct the minutiae pairs with the incorporation of partial discrete Fourier transform in order to generate discriminating features for matching. The scheme however, only shows good accuracy performance on DB1 of FVC2002, while achieving a relatively low accuracy in other datasets. In another different work \cite{RN41}, the authors used a partial Hadamard transform approach for the design of cancelable templates. The stochastic distance between binary vectors is theoretically proven to be preserved after transformation. Even though the achieved accuracy is better than the previous scheme, it still lacks a profound distance preservation analysis.
 
Recently, Trivedi et al. \cite{RN40} proposed a non-invertible approach based on information extracted from the Delaunay triangulation of the minutiae points. A random binary sting is utilized with a user specific key for cancelability. However, the scheme is evaluated on a single dataset only, which of course is insufficient for effective validation of the accuracy performance and other cancelable properties of the scheme. Another novel design of cancelable templates using randomized non-negative least squares is performed by Kho et al. \cite{RN34}. The authors formulated a unique PR-NNLS descriptor for non-invertible transformation but instead of applying it directly on the minutiae descriptor, they chose to apply it on the PLS descriptor dictionary in order to retain the accuracy. However, its full non-invertibility still cannot be ascertained because there is no analysis conducted on the distance preservation. In a rather similar approach by Shahzad et al \cite{RN39}, authors explored the concept of window-shift-XOR and discrete wavelet transform to avert ARM. Some similarity-based attacks are evaluated on their dual protection mechanism to validate the non-invertibility of the scheme but without an analysis.

The new prevailing problem applicable to most cancelable biometric systems is their inability to achieve full non-invertibility due to the fact that the transformation techniques often used are not preserving the distance space between original features and their transformed versions. However, the best possible ways of overcoming these issues are 1) to avoid matching in the transformed domain, instead perform matching in an encrypted domain (which of course requires the integration of an encryption mechanism); 2) perform matching in an entirely different dataset that does not require a lot of attempts for same user verification; 3) develop schemes that take full cognizance of the non-invertibility of the system, e.g., by incorporating additional protection mechanisms to secure the parameters \cite{RN45}. Gomez et al. \cite{RN29} gave a comprehensive survey on the inversion of biometric templates with possible ways of surmounting the weakness associated with the related attacks. One of such attacks is launched through perceptron learning \cite{RN26}. A deep secure quantization approach\cite{RN25}, genetic algorithm \cite{patrick2013}, and constraint optimization approach \cite{wang2021} are some of the recent works aimed at preventing such attacks. Another survey by Dong et al.\cite{dong2020} analyzes the distance preserving properties in cancelable biometrics. They also digressed into the security risks in relation to pre-image attacks, and proposed a framework for countering such attacks. Their previous works on random projection \cite{jin2014} \cite{andrew2006}, index-of-max hashing \cite{zjin2018}, and kernalized hashing \cite{kiran2018} are however, susceptible to similarity-based attacks. A cryptanalysis conducted by Ghamman et al. \cite{RN28} also proves the vulnerability of \cite{zjin2018}.

\section{Proposed scheme}

First, the minutiae feature descriptor, popularly known as Minutiae Cylinder Code (MCC)\cite{RN24} is employed to extract fingerprint vector features, and a fixed length representation technique using Kernel Principle Component Analysis (KPCA)\cite{RN33} to convert the extracted feature descriptors into fixed-length vectors is applied. We choose to use MCC feature descriptor due to its invariance property and its capability of local minutiae matching, which leads to high performance accuracy during authentication. The accuracy is also retained after feature transformation using KPCA, which has also proven better performance and accuracy preservation than other descriptors \cite{helala2018} \cite{alessandra2013}.

\subsection{Feature Extraction with MCC}

This is done by capturing the directional and spatial relationship between the reference minutiae 
${m_{r} =\left\{p_{r} ,q_{r} ,\theta _{r} \right\}}$. For $p_{r} $ and $q_{r} $ denotes the minutiae point location, while $\theta _{r} $ represents the minutiae orientation. Their corresponding aligned neighbor minutiae ${m^{t} =\left\{p_{i}^{t}, {q_{i}{t}, \theta '_{i}{t}} |i=1,...,N-1\right\}}$ is within the selected pre-defined radius R. The extraction technique mainly relies on a robust discretization of the neighborhood of each minutia into a 3D cell-based structure known as the cylinder, as each cylinder is discretized into a number of cells. Hence, a numerical value is computed from each cell based on the likelihood of finding minutiae that are closer to the cell with respect to location and direction. MCC is a prominent fingerprint recognition technique based on minutiae template. Due to its cylinder invariance, fixed-length binary string, and bit-oriented coding, it provides us with the perfect suitability for achieving non-invertibility, cancelability and unlinkability of our proposed template protection technique.\\[6pt]
\textbf{Generate the feature vector:}
In order to reduce the dimension of the binary strings derived from MCC, KPCA is employed for such purpose. To achieve this, MCC descriptor variable size is converted into a fixed-length vector representation. Given ${\Phi  = \left\{ {{\Phi ^l}(i)|i = 1,...{N_l}} \right\}}$ as the set of MCC training samples, where ${N_l}$ is the total number of ${\Phi ^l}$, the kernel matrix of the fixed features ${F_k} \in \mathbb{Q}{^{{v_l} \times {v_l}}}$ is computed with its resultant function as given in Eq. 1 below. 
\begin{equation}
\begin{array}{l}
{F_f}\left( {i,j} \right) = f\left( {{\Phi ^i}\left( i \right),{\Phi ^i}\left( j \right)} \right) = \\
\exp \left( { - 1{{\left( {1 - {D_{{s_{MCC}}}}\left( {i,j} \right)} \right)}^2}/2{\sigma ^2}} \right)
\end{array}
\end{equation}
 where ${D_{{s_{MCC}}}}$ denotes the MCC dissimilarity measure, given ${\sigma ^2}$ as the speed factor. The eigenvectors of KPCA projection matrix ${P_r} \in \mathbb{Q}{^{{v_l} \times d}}$ is derived as in \cite{scholkopf1998}, where $d$ represents the output dimension of a considerate desired projection. Then, a query instance of the MCC given as ${\Phi ^q}$ is used to match with all the training samples ${\Phi ^l}(i)$, such that $1 \le i \le {N_l}$. Afterwards, ${N_l}$ is concatenated with ${v_i}$ as the matching scores ${v_i}: \leftarrow sim\left( {{\Phi ^q},{\Phi ^l}(i)} \right)$, in order to generate a vector $v \in \mathbb{Q}{^{{v_l}}}$. $v$ can then be transformed using the initial kernel function to have ${F(v) = exp\left( { - 1{{\left( {1 - v} \right)}^2}/2{\sigma ^2}} \right)}$. Hence, the fingerprint vector is generated through $\omega  = v'T \in \mathbb{Q}{^d}$, where ${v' = F(v)}$, for $v' \Rightarrow v \in \mathbb{Q}{^{{N_l}}}$, and $T \in \mathbb{Q} {^{{N_l} \times d}}$.

\subsection{Feature vector permutation}

The generated fixed-length vectors from KPCA are used as the input to our protection mechanism. For each of the derived vectors $v \in \mathbb{Q}{^{{N_l}}}$ and $T \in \mathbb{Q}{^{{N_l} \times d}}$, we utilized an instance of LSH \cite{charikar2002} from the function $h(.)$ to generate ${h_i}$ together with a permutation set $\left[ {{P_\theta }} \right]$. Lets assume the parent fingerprint feature vector $\omega  = v'T \in \mathbb{Q}{^d}$ and an $n$ independent LSH family hash function denoted by ${h_i}(\omega ) \in [1,b]|i = 1,...,n$, where each hash function constitutes a product function. Therefore, for each ${h_i}(\omega )$, $p$ independent random permutation seeds are generated to make up the permutation set $\left[ {{P_\theta }} \right]$. Consequently, $\omega$ is permuted so as to get the permuted vectors ${\bar \omega _k} = {\mathbb{Q}^d}|k = 1,...,p$. Next, a hash sample ${h_{({{\bar \omega }_k}i)}}$ is formed by performing $n$ times permutation of randomly independent enrolled vectors. Hence, ${h_{({{\bar \omega }_k}i)}}$ yields $n$ number of permuted feature vector ${h_{({{\bar \omega }_k}i)}} \in \mathbb{Q}{^{{v_l}}}|i = 1,...,n.$ \\[6pt]
\textbf{Sampling bit location:}
The hash generation procedure can be denoted by the function $\mathbb{F}$ which constitutes of the randomly sampled bit location $\xi$ of each permuted hash vector ${h_{({{\bar \omega }_k}i)}}$. Thus, each hash sample can be denoted by Eq. 2 below. 
\begin{equation}  
{\rm {\mathbb{F}}}\left(\xi \right):{\rm {\mathbb{Q}}}^{v_{l} } \to h_{(\bar{\omega }_{k} i)} \left({\rm {\mathbb{Q}}}_{i} \right)|\left\{i=1,2,...,n\right\} 
\end{equation} 
We extract the bits ${\theta _{bits}}$ corresponding to each marked position $b$ of $\xi _{b} \left(\theta _{(h_{i,n}^{'} )_{bits} } \right)$ from the entire hash components $H_{C_{(i,n)} } $, where $h_{i}^{'} $ denotes the permuted hash vector deduced from ${h_{({{\bar \omega }_k}i)}}$. Subsequently, the bit position of each vector element $i = \left\{ {1,2,...,n} \right\}$ from the vector component ${C_{(i,n)}}$ is retained. Then its equivalent is concatenated within the hash component ${H_{{C_{(i,n)}}}}$. Thus, we have $\xi _{b} \left(\theta _{(h_{i,n}^{'} )_{bits} } \right)||H_{C_{(i,n)} } $.

\subsection{Non-invertible hash component}
Even though the randomized permutation of the enrolled vectors in the previous step can provide us with sufficient security, we still consider storing ${C_{(i,n)}}$ in the plain as insecure because an attacker can make use of the vector components to effectively infer to the feature vectors in each of the components available in the marked position. To avoid such scenario from occurring, a security parameter ${S_k}$ is introduced such that ${S_k}:{L_{ow}} \le {S_k} \le {H_{igh}}$ for all permuted vector components. ${L_{ow}}$ and ${H_{igh}}$ represent the predefined lowest and highest values of the given threshold, respectively. This threshold is also tied to the accuracy preservation of the scheme. Hence, eliminating or limiting the trade-off impact between accuracy and non-invertibility. Consequently, ${S_k}$ can be incorporated into each vector component with respect to their bit location using Eq. 3 below. 
\begin{equation}
S_{k} \left[{\rm {\mathbb{Q}}}_{C_{(i,n)} }^{'} \right]=\xi _{b} \left(\theta _{(h_{i,n}^{'} )_{bits} } \right)||H_{C_{(i,n)} }  
\end{equation} 
Next, we perform a random search via a shift-max order process on the entire hash component ${H_{{C_{(i,n)}}}}$ by taking the index of the maximum value from the 1st two elements (in which one of them must be the highest in the entire hash component). We term this as the shift-max value order scenario. Each component ${H_{{C_{(i,n)}}}}$ consists of $m$ number of elements with a fixed window size $k$. Thus, ${H_{{C_{(i,n)}}}}$ can be represented as ${H_{{C_{(i,n)}}}} = \left[ {H_{{1_{(i,n)}}}^C,H_{{2_{(i,n)}}}^C,...,H_{{m_{(i,n)}}}^C} \right]$ and $m = l/k$, where $l$ denotes the length of the vector component. In the first order shifting, each $H_{{1_{(i,n)}}}^C$ is shifted in a sequentially cyclical order in each of the hash component ${H_{{C_{(i,n)}}}}$, until the last element $H_{{m_{(i,n)}}}^C$. The entire shifting procedure can be represented by the matrix in Eq. 4. 
\begin{align}
\begin{tiny}
\begin{array}{l}
{M_{shift}} = \left[ {\begin{array}{*{20}{c}}
{H_{{1_{(i,n)}}}^C}\\
{H_{{2_{(i,n)}}}^C}\\
 \vdots \\
{H_{{m_{(i,n)}}}^C}
\end{array}} \right] = \\
{\left[ {\begin{array}{*{20}{c}}
{H_{{1_{(i,n)}}}^C\left( a \right)}\\
{H_{{2_{(i,n)}}}^C\left( a \right)}\\
 \vdots \\
{H_{{m_{(i,n)}}}^C(a)}
\end{array}{\rm{    }}\begin{array}{*{20}{c}}
{H_{{1_{(i,n)}}}^C\left( b \right){\rm{ }} \cdots }\\
{H_{{2_{(i,n)}}}^C\left( b \right){\rm{ }} \cdots }\\
 \vdots \\
{H_{{m_{(i,n)}}}^C\left( b \right){\rm{ }} \cdots }
\end{array}\begin{array}{*{20}{c}}
{H_{{1_{(i,n)}}}^C\left( k \right)}\\
{H_{{2_{(i,n)}}}^C\left( k \right)}\\
 \vdots \\
{H_{{m_{(i,n)}}}^C\left( k \right)}
\end{array}} \right]_{m \times k}}
\end{array}
\end{tiny}
\end{align}
where $\left[ {H_{{1_{(i,n)}}}^C\left( a \right){\rm{,    }}H_{{1_{(i,n)}}}^C\left( b \right){\rm{ }} \cdots {\rm{ }}H_{{1_{(i,n)}}}^C\left( k \right)} \right]$ to $\left[ {H_{{m_{(i,n)}}}^C\left( a \right){\rm{,   }}H_{{m_{(i,n)}}}^C\left( b \right){\rm{ }} \cdots {\rm{ }}H_{{m_{(i,n)}}}^C\left( k \right)} \right]$ are the first and last components constituting the entire elements of a given  $k-size$ window, while $m \times k$ denote the size of the entire ${M_{shift}}$ matrix. 

After selecting the highest value from the 1st two elements $\left[ {H_{{1_{(1,n)}}}^C\left( a \right),{\rm{ }}H_{{1_{(1,n)}}}^C\left( b \right)} \right]$ of the 1st component $\left[ {H_{{1_{(i,n)}}}^C} \right]$, it is then passed on to the second component $\left[ {H_{{2_{(i,n)}}}^C} \right]$, until the entire component is searched whilst preserving the bits position of the selected elements (i.e., shifting bits corresponding to each element in the vector components according to the matrix order). From the second component, the highest value of $h_{i}^{'} $ from each component is thus passed to the hash bin ${H_{bin}}$ for final hash construction. As such, the randomized permutation, security parameter ${S_k}$, coupled with the shift-max value procedure, all contribute to the non-invertibility of the templates. Fig. 1 illustrate the entire hash generation procedure at both enrollment and querying stages.\\[3pt]                                                                                                                                            
\begin{figure*}
\setlength\abovecaptionskip{0.2\baselineskip}
\setlength\belowcaptionskip{-2.5pt}
\centering
\includegraphics[width=1.0\linewidth]{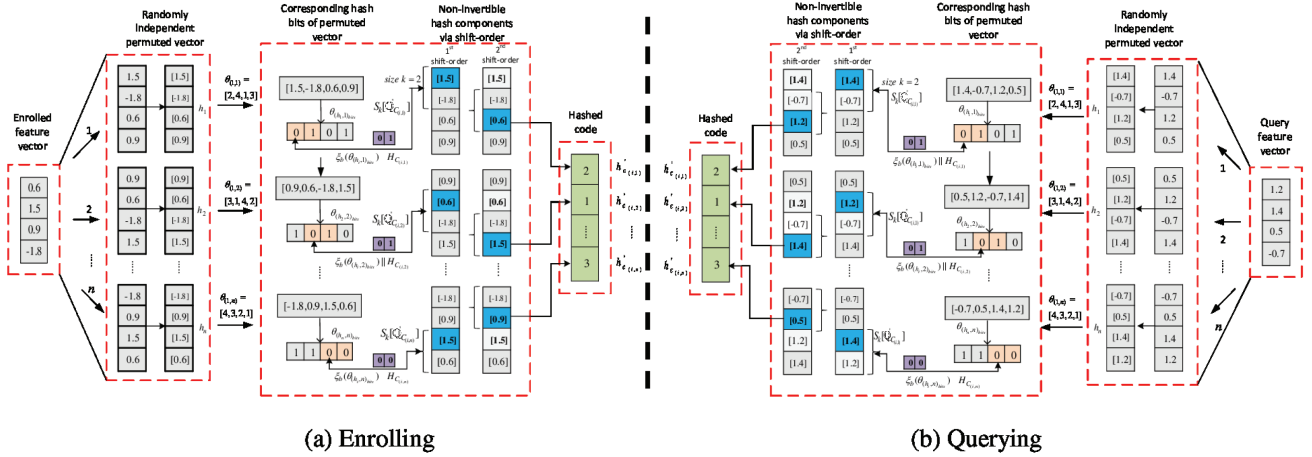}
\caption{Illustration of feature vector enrolling and querying for hash code generation}
\end{figure*} 
\textbf{Final hash construction:}
In order to construct the final hash code, we take the index of the highest value from the bin such that for each $h_{i}^{'} \in H_{bin} $, we have;
\begin{equation}
\xi _{b} \left(\theta _{(h_{i,n}^{'} )} \right)=\max _{j=1,...,n} \left\langle h_{i}^{'} ,y_{i}^{'} \right\rangle |i=1,...,n 
\end{equation} 
for $h_{i}^{'} \in H_{bin} $ and $y_{i}^{'} \in H_{bin} $, where $H_{bin} $ denotes the hash bin, and $y_{i}^{'} $ denote the index of the highest value in the bin. The highest value is then recorded in the first $k-size$ window and its index value is stored. Therefore, we will have $\left[\theta _{h_{(i,n)}^{'} } \right]\in \left[1,k\right]{\rm \; for\; }i=1,...,n$. We repeat stage 2 and 3 on different hash components using different permutation sets, and construct the final hash code $h_{{C_{(i,n)}}}^{''}|i = 1,...,n$. Note that every single hash component ${H_{{C_{(i,n)}}}}$ constitutes of the hash code $h_C^{''}$ that emanates from the permuted feature vector $h_{(\bar{\omega }_{k} i)} \in {\rm {\mathbb{Q}}}^{v_{l} } $. Hence, \begin{equation} 
h_{C}^{''} =h_{(\bar{\omega }_{k} i)} \in {\rm {\mathbb{Q}}}^{v_{l} } [1,b]|i=1,...,n 
\end{equation} 

\textbf{Similarity score:}
A similarity score is used to evaluate the comparison between the reference (enrolled) and probe (queried) hash codes of the proposed scheme. Since LSH instance is used to ascertain the similarity of permuted vector features within the inherent set of ordered elements, whereby their hash is directly dependent on their relative distances, we therefore cannot opt for traditional similarity measures that does not support such ordering. Instead, the global similarity is estimated between the enrolled and queried hash codes by computing the number of elements within each component that matches with the two sets. Each set here correspond to the correlation order of each index of the maximum value over $k-size$ window. In order to enhance security, matching between the enrolled and queried vectors is still conducted in the transformed domain, but by taking into cognizance the similarity between feature vectors with respect to the shifting process, which might be argued of being prone to similarity-based attacks (Our future work will give an in-depth analysis on this).\\For each hash code $h_{{C_{(i,n)}}}^{''}$ and $h_{{C_{(j,n)}}}^{''}$, let $h_{{C_{(i,1)}}}^{''},h_{{C_{(i,2)}}}^{''},...,h_{{C_{(i,n)}}}^{''}$ and $h_{{C_{(j,1)}}}^{''},h_{{C_{(j,2)}}}^{''},...,h_{{C_{(j,n)}}}^{''}$ be the enrolled and queried instances of the non-invertible hash components, respectively. Thus, the two sets of hash components from $h_{{C_{(i,n)}}}^{''}$ and $h_{{C_{(j,n)}}}^{''}$ can be given as: 
\begin{equation}
E\left[ {h_{{C_{(i,n)}}}^{''}} \right] = h_{{C_{(i,1)}}}^{''},h_{{C_{(i,2)}}}^{''},...,h_{{C_{(i,n)}}}^{''}|i = 1,...,n
\end{equation}
\begin{equation}
Q\left[ {h_{{C_{(j,n)}}}^{''}} \right] = h_{{C_{(j,1)}}}^{''},h_{{C_{(j,2)}}}^{''},...,h_{{C_{(j,n)}}}^{''}|j = 1,...,n
\end{equation}
Suppose the collision probability of the two hash codes is $CP\left\{ {E\left[ {h_{{C_{(i,n)}}}^{''}} \right]} \right.,\left. {Q\left[ {h_{{C_{(j,n)}}}^{''}} \right]} \right\}$. Therefore,
\begin{equation}
\begin{array}{l}
{\mathbb{C}_P} = \left[ {E\left[ {h_{{C_{(i,n)}}}^{''}} \right] = \left. {Q\left[ {h_{{C_{(j,n)}}}^{''}} \right]} \right]} \right. \Rightarrow \\
CP\left\{ {E\left[ {h_{{C_{(i,n)}}}^{''}} \right]} \right.,\left. {Q\left[ {h_{{C_{(j,n)}}}^{''}} \right]} \right\}{\rm{  for }}i = 1,...,n
\end{array}
\end{equation} 
Hence, the similarity between $E\left[ {h_{{C_{(i,n)}}}^{''}} \right]$ and $Q\left[h_{C_{(j,n)} }^{''} \right]$ indicates a high probability, while their dissimilarity indicates a less collision probability. Consequently, a normalized similarity correlation score is derived by dividing the number of matches between the two components with their set cardinality. The similarity correlation between the two hash codes can be given as: 
\begin{equation} 
S_{C} =\frac{\sum _{b=1}^{n}\left|E\left[h_{C_{(i,n)} }^{''} \right]^{r} =Q\left[h_{C_{(j,n)} }^{''} \right]^{r} \right|/k }{V_{l}^{'} }  
\end{equation} 
where $V_{l}^{'} =V_{l} /k$ for each ${V_l}$ denoting a single vector element from the vector component, $r$ is the correlation order of each index, and $b$ is the index of maximum value over $k-size$ window. This $b$ can be selected via a $k-1$ entries that are less than $b$ in both the enrolled and queried hash codes, which helps determine the exact similarity between two hash codes and consequently assist with the selection of the maximum element within a hash component. 

\section{Experiments}
In this section, we performed an extensive evaluation of the proposed template protection scheme over all six public fingerprint datasets which includes, FVC2002 (DB1, DB2, and DB3), as well as FVC2004 (DB1, DB2, and DB3). Each dataset contains 800 fingerprints of varying quality collected from 100 users with 8 impressions per user. Verifinger SDK is utilized for the extraction of ISO-compliant minutiae information which was further feed to MCC. Four major parameters including security threshold ${S_{key}}$, window size $k$, Hashvecs ${H_{{C_{(i,n)}}}}$, and number of elements in each vector $Nu{m_{perm}}$ are assigned specific values with respect to analyzing the accuracy of the proposed scheme. Also, the security and privacy analysis including non-invertibility, cancelability/revocability, and unlinkability, were all carried out in order to evaluate the efficacy of the proposed scheme. 

Different performance measures are used for the evaluation. These are the False Match Rate (FMR), False Non-Match Rate (FNMR), and Equal Error Rate (EER). FMR results from the decision threshold of a biometric system to misinterpret the probability of two different impressions from two fingers as an impression of one finger, while FNMR results from the decision of the system to misinterpret the probability of two different impressions from one finger as an impression of two fingers. The common value between FMR and FNMR is the ERR, which determines the accuracy of the scheme. (i.e., the error rate when FMR = FNMR). Likewise, the Genuine Match Rate (GMR) is akin to the FMR when GMR + FMR = 1. We employed the original FVC protocol for all evaluations and analysis.

\begin{figure*}[h]
\begin{center}
\setlength\abovecaptionskip{0.2\baselineskip}
\setlength\belowcaptionskip{-2.5pt}
\includegraphics[width=1.0\linewidth]{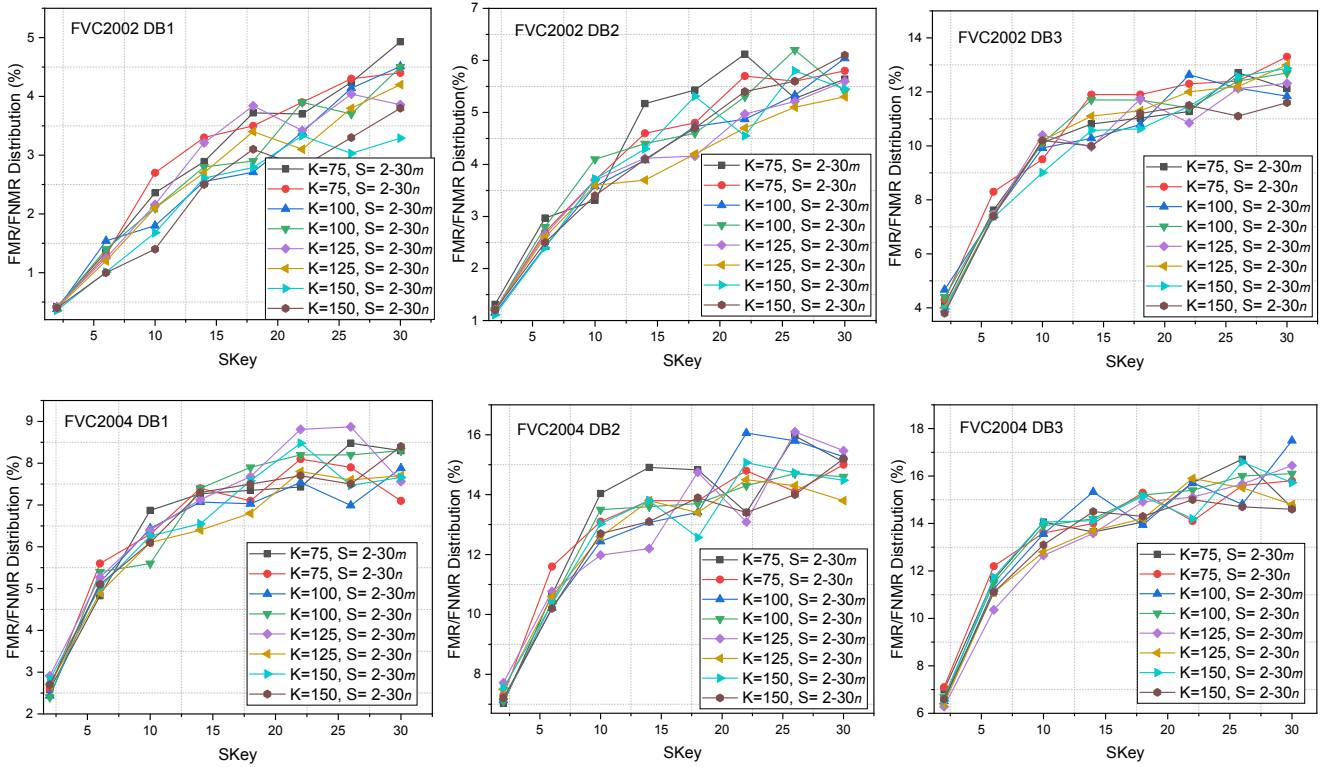}
\caption{Effect of security threshold and window size w.r.t. FMR and FMNR}
\end{center}
\end{figure*}

\begin{table*}[h]
\setlength\abovecaptionskip{0.2\baselineskip}
\setlength\belowcaptionskip{-2.5pt}
\begin{center}
\begin{tabular}{cllllllllllll} \hline
\multicolumn{1}{c}{}                                                                       & \multicolumn{2}{c}{\textbf{FVC2002DB1}} & \multicolumn{2}{c}{\textbf{FVC2002DB2}} & \multicolumn{2}{c}{\textbf{FVC2002DB3}} & \multicolumn{2}{c}{\textbf{FVC2004DB1}} & \multicolumn{2}{c}{\textbf{FVC2004DB2}} & \multicolumn{2}{c}{\textbf{FVC2004DB3}} \\
\multicolumn{1}{c}{\multirow{-2}{*}{\textbf{${S_{key}}$}}}                                        & \multicolumn{1}{l}{GMR}     & FMR    & \multicolumn{1}{l}{GMR}     & FMR    & \multicolumn{1}{l}{GMR}     & FMR    & \multicolumn{1}{l}{GMR}     & FMR    & \multicolumn{1}{l}{GMR}     & FMR    & \multicolumn{1}{l}{GMR}     & FMR    \\ \hline
2    & 99.61          & 0.38          & 98.80          & 1.15          & 96.00          & 3.94          & 97.60          & 2.42          & 92.70          & 7.41          & 93.40          & 6.95          \\
6    & 99.00          & 0.95          & 97.40          & 2.85          & 91.60          & 8.10          & 94.80          & 5.52          & 88.60          & 11.01         & 89.00          & 11.41         \\
10   & 97.50          & 2.18          & 96.10          & 3.64          & 90.40          & 8.59          & 93.60          & 7.01          & 87.60          & 13.25         & 85.80          & 13.09         \\
14   & 97.00          & 2.59          & 94.80          & 4.57          & 88.70          & 10.61         & 92.50          & 6.55          & 86.30          & 13.41         & 84.70          & 14.26         \\
18   & 96.30          & 3.05          & 94.70          & 5.11          & 89.20          & 11.47         & 93.20          & 7.62          & 86.60          & 13.88         & 84.20          & 14.55         \\
22   & 96.10          & 3.64          & 94.30          & 5.03          & 88.00          & 12.83         & 91.90          & 7.17          & 85.70          & 14.34         & 85.20          & 16.51         \\
26   & 96.10          & 3.33          & 94.90          & 5.31          & 88.60          & 12.73         & 92.30          & 8.61          & 86.40          & 15.64         & 84.30          & 14.61         \\
30   & 96.20          & 4.36          & 94.10          & 5.37          & 87.40          & 12.79         & 91.80          & 7.70          & 84.90          & 14.91         & 83.90          & 15.13        \\ \hline
\end{tabular}
\end{center}
\caption{Performance accuracy of GMR against FMR w.r.t. the security threshold.}
\end{table*}
\subsection{Effect of security threshold ${S_{key}}$ and window size $k$}
In this analysis, the number of hash vector ${H_{{C_{(i,n)}}}}$ in each component  ${{C_{(i,n)}}}$ (component size) is set at 10, while the number of vector elements is set at 500. The affected parameters i.e., window size $k$ and security threshold ${S_{key}}$ are set within the specified range of $k$ = 75, 100, 125, 150, and ${S_{key}}$ = 2, 6, 10, 14, 18, 22, 26, 30, respectively. Fig. 2. shows the overall plots analysis against FMR and FNMR with the initials $m$ and $n$ as depicted on their distributions $(K,  S)$, respectively. All six datasets of FVC2002 and FVC2004 are utilized for the evaluation. It can be seen that lower ${S_{key}}$  and higher $k$ generally exhibits better accuracy as compared to higher ${S_{key}}$  and lower $k$. This proves the sensitivity of the generated hash codes w.r.t the defined security threshold. Also, DB1 \& DB2 of FVC2002, and DB1 of FVC2004 depicts a better performance than the other datasets.

\begin{table*}[t]
\setlength\abovecaptionskip{0.2\baselineskip}
\setlength\belowcaptionskip{-2.5pt}
\begin{center}
\begin{tabular}{@{}lcccccc@{}}
\toprule
\textbf{Techniques}                        & \textbf{\begin{tabular}[c]{@{}c@{}}FVC2002\\ DB1\end{tabular}} & \textbf{\begin{tabular}[c]{@{}c@{}}FVC2002\\ DB2\end{tabular}} & \textbf{\begin{tabular}[c]{@{}c@{}}FVC2002\\ DB3\end{tabular}} & \textbf{\begin{tabular}[c]{@{}c@{}}FVC2004\\ DB1\end{tabular}} & \textbf{\begin{tabular}[c]{@{}c@{}}FVC2004\\ DB2\end{tabular}} & \textbf{\begin{tabular}[c]{@{}c@{}}FVC2004\\ DB3\end{tabular}} \\ \hline
\multicolumn{7}{c}{\cellcolor[HTML]{EFEFEF}\textbf{With Template Protection Mechanism}}                                                                                                                                                                                                                                                                                                                                                          \\
$2{\rm{P MC}}{{\rm{C}}_{128,{\rm{ 128}}}}$ \cite{RN27}                                & 1.88                                                           & 0.99                                                           & 1.24                                                           & -                                                              & 4.84                                                           & -                                                              \\
$2{\rm{P MC}}{{\rm{C}}_{64,{\rm{ 64}}}}$ \cite{RN27}                                  & 3.33                                                           & 1.76                                                           & 7.78                                                           & -                                                              & 6.62                                                           & -                                                              \\
$2{\rm{P MC}}{{\rm{C}}_{16,{\rm{ 16}}}}$ \cite{RN27}                                  & 12.39                                                          & 10.16                                                          & 19.33                                                          & -                                                              & 17.70                                                          & -                                                              \\
Bloom filter with fingerprint \cite{narishige2015}             & 2.3                                                            & 1.8                                                            & 6.6                                                            & 13.4                                                           & 8.1                                                            & 9.7                                                            \\
PR-NNLS $(\gamma  = {N_c} + 150)$ \cite{RN34}                                  & 2.28                                                           & 1.25                                                           & 6.4                                                            & -                                                              & 7                                                              & -                                                              \\
PR-NNLS $(\gamma  = {N_c} + 20)$ \cite{RN34}                                  & 2.48                                                           & 1.51                                                           & 7.03                                                           & -                                                              & 7.44                                                           & -                                                              \\
IoM Hashing \cite{zjin2018}                               & 0.43                                                           & 2.10                                                           & 6.60                                                           & 4.51                                                           & 8.02                                                           & 8.46                                                           \\
Fractal Coding Hashing \cite{sani2020}                    & 0.36                                                           & 0.54                                                           & 2.395                                                          & 2.348                                                          & 5.925                                                          & 2.365                                                          \\
\textbf{Proposed (Protected)}              & \textbf{0.38}                                                  & \textbf{1.16}                                                  & \textbf{3.88}                                                  & \textbf{2.42}                                                  & \textbf{7.12}                                                  & \textbf{6.34}                                                  \\
\multicolumn{7}{c}{\cellcolor[HTML]{EFEFEF}\textbf{Without Template Protection Mechanism}}                                                                                                                                                                                                                                                                                                                                                       \\
\textbf{MCC (Original)}                    & \textbf{0.60}                                                  & \textbf{0.59}                                                  & \textbf{3.91}                                                  & \textbf{3.79}                                                  & \textbf{5.22}                                                  & \textbf{3.82}                                                  \\
\textbf{Fixed-length Vector (Transformed)} & \textbf{0.20}                                                  & \textbf{0.19}                                                  & \textbf{2.30}                                                  & \textbf{4.70}                                                  & \textbf{3.13}                                                  & \textbf{2.80}                                                  \\
TKLSH-DQ \cite{RN33}                                  & 3.76                                                           & 4.94                                                           & 11.58                                                          & 17.11                                                          & 12.19                                                          & 14.50                                                          \\
TKPCA-DQ \cite{RN33}                                  & 4.03                                                           & 5.04                                                           & 11.86                                                          & 17.39                                                          & 10.95                                                          & 13.49                                                          \\ \hline
\end{tabular}
\end{center}
\caption{Accuracy performance comparison between the proposed and other state-of-the-art schemes}
\end{table*}
                                                                                                                                            
\begin{figure}[!h]
    \setlength\abovecaptionskip{0.2\baselineskip}
    \setlength\belowcaptionskip{-2.5pt}
    \centering
    \includegraphics[width=3.1in]{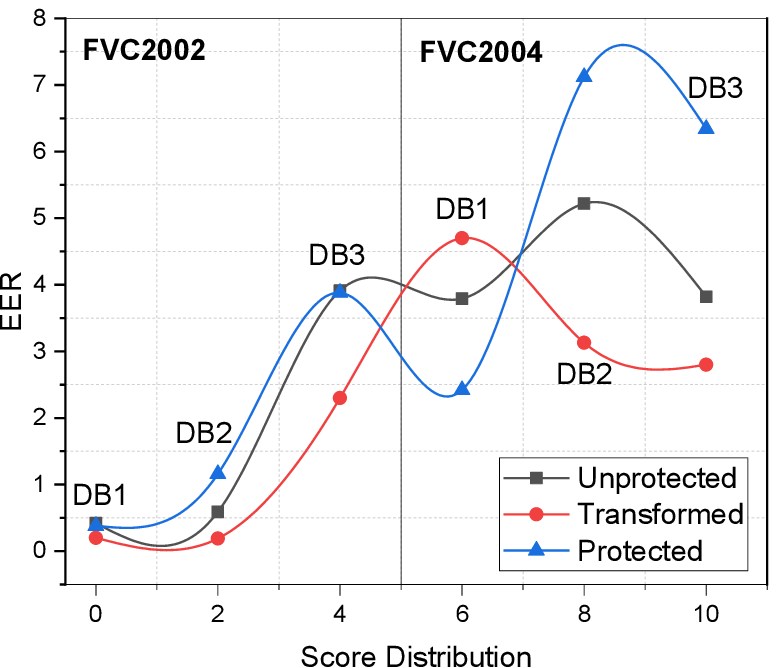}
    \caption{Effect of security threshold and window size w.r.t. FMR and FMNR}
\end{figure} 
\floatsep
\belowcaptionskip                                                                                                                                            
\subsection{Performance Accuracy and Comparison with other state-of-the-arts}
In addition to the evaluations with respect to the effect of different parameters, we also analyze the matching performance of the scheme under lost-key scenario using the FVC protocol, in relation to the GMR and FMR. All six datasets are used for the experiment by keeping the window size $k$ at 100, $Nu{m_{perm}}$ at 500. The security threshold ${S_{key}}$ is however adjusted from 2 to 30 according to the previous experimentation while keeping the hash vector at 5. A detailed description of the evaluation is given in Table 1 below. It can easily be judged from the Table that the best performance in all datasets is achieved when the security threshold  is set at 2. To further analyze the performance accuracy of the scheme, we compare our approach with other well-known template protection schemes. By setting the security threshold ${S_{key}}$ = 2 and Hashvec ${H_{{C_{(i,n)}}}}$ = 10, we record the EER of the proposed method under original FVC protocol (with window size $k$ = 150 for all FVC2002 datasets, $k$ = 100 for DB1 and DB2 of FVC2004, and $k$ = 125 for DB3 of FVC 2004). Table 2 shows the performance comparison between the proposed method and other state-of-the-art existing cancelable templates schemes with protection and their corresponding transformed version before the incorporation of protection mechanism.\\
It can be seen from Table 2 that the proposed scheme outperforms all other techniques except for fractal coding-based hashing. However, comparing 2PMCC at its best $\left( {2{\rm{P MC}}{{\rm{C}}_{128,{\rm{ 128}}}}} \right)$, moderate $\left( {2{\rm{P MC}}{{\rm{C}}_{64,{\rm{ 64}}}}} \right)$, and worst $\left( {2{\rm{P MC}}{{\rm{C}}_{16,{\rm{ 16}}}}} \right)$ parameters, we can notice that our scheme still outperforms $2{\rm{P MC}}{{\rm{C}}_{128,{\rm{ 128}}}}$ (best) on FVC2002 DB2, and $2{\rm{P MC}}{{\rm{C}}_{64,{\rm{ 64}}}}$ (moderate), and $2{\rm{P MC}}{{\rm{C}}_{16,{\rm{ 16}}}}$ (worst) with a huge margin. For the best $(\gamma  = {N_c} + 150)$ and worst $(\gamma  = {N_c} + 20)$  performances of PR-NNLS based scheme, it shows a slightly better performance on DB2 of FVC2004 at $(\gamma  = {N_c} + 150)$. As for techniques without incorporation of template protection, the transformed fixed length vector outperforms the original MCC with a huge margin as compared to the proposed protected version (due to the slight manifestation of trade off between accuracy and security even with a moderate Hashvec ${H_{{C_{(i,n)}}}}$ value. Meanwhile, the protected version also outperforms the original MCC. The EER of all 3 approaches from same features is depicted in Fig. 3, with MCC as unprotected, fixed-length vector as transformed, and the proposed scheme as protected. It can be seen that the protected scheme generally outperforms both unprotected and transformed version in BD1 of FVC2004, whilst being outperformed by its counterparts in DB2 and DB3 of same database. This certainly has to do with the low identifiable correlation within different elements in the components of these datasets, which in same vain emanates from the distorted or low-quality images in the datasets.  In FVC2002 however, the protected version performs better than its unprotected counterpart in DB1 and DB2. But is slightly outperformed by the transformed version, which of course has to do with the incorporation of security mechanisms. Let’s also not forget that the rigorous search for a maximum value between the 1st two elements of each components has a role to play in reducing the accuracy performance of the protected templates. Overall, $2{\rm{P MC}}{{\rm{C}}_{16,{\rm{ 16}}}}$ shows the worst performance on schemes incorporated with a template protection mechanism, TKPCA-DQ depicts the worse performance on schemes without a template protection mechanism, while the fixed length (transformed) version performs best among the three techniques from same features. 
Also, it is worth mentioning that due to the randomness (caused by random vector permutation) and iterative reshuffling (caused by shifting procedure) of the mechanism, a difference of 0.1 to 0.5 is noticed at each specific feature reconstruction of the protected template when a loop iteration without pre-allocation is allowed. This certainly confirms the sensitivity of the hash code and the entire security mechanism of the scheme.
\subsection{Revocability}
Revocability is an important property of cancelable biometrics which pertains to the security of a biometric system. When the original stored biometric template is compromised, it is expected by the cancelable system to re-generate uncorrelated templates from same fingerprint. Thus, further ensuring that the reissued templates from same fingerprint are not correlated despite being generated from the same biometric data. To test the revocability of the proposed scheme, transformed templates are generated from first impression of each finger using different randomly generated keys. Same process is repeatedly applied to DB1 of FVC2002 by matching the pseudo-imposter templates against its imposter and genuine counterparts. The mean for both pseudo-impostor and impostor distributions are set at 0.35. Fig. 4 shows the distributions of pseudo-impostor, impostor, and genuine scores of the dataset. We can see from the figure that the distributions of pseudo-impostor and impostor overlap, while being completely unrelated to the genuine distributions. This signifies that the reissued version of the fingerprint templates are uncorrelated despite coming from same source.
\begin{figure}[!h]
    \setlength\abovecaptionskip{0.2\baselineskip}
    \setlength\belowcaptionskip{-2.5pt}
    \centering
    \includegraphics[width=3.1in]{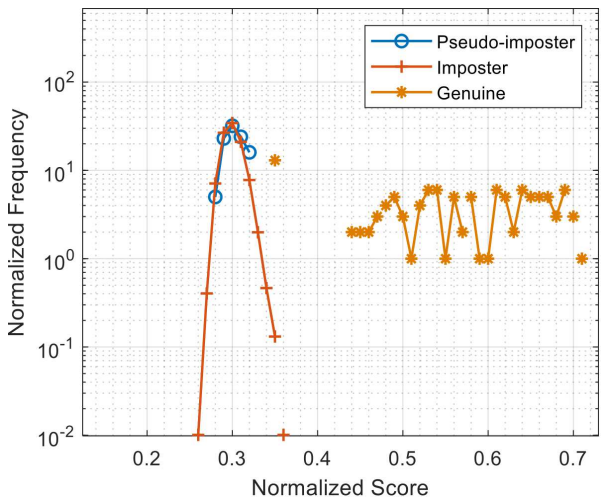}
    \caption{Revocability analysis}
\end{figure}\\
\floatsep
\belowcaptionskip
\subsection{Unlinkability}
Unlinkability is another important property of cancelable biometrics which relates to the privacy of the biometric data. To measure the unlinkability of a cancelable biometric system, it is required that the protected template should not cross-match with the same corresponding unprotected fingerprint on different platforms and/or applications. Two major indicators are used for the evaluation of unlinkability. These are mated and non-mated score distributions. According to the framework \cite{RN30}, mated and non-mated sample score distributions measure the system’s unlinkability at both local and global level. Calculating the mated sample scores is mainly performed by comparing the templates that are extracted from fingers with same impression using different keys, while non-mated scores are computed by comparing templates that are extracted from fingers with different impressions using different keys. Unlinkability analysis of the scheme is conducted on DB1 of FVC2002 dataset. The same mean value used in revocability is also maintained here. The score distributions between mated and non-mated samples can be seen in Fig. 5, which revealed a significant overlap. Thus, indicating the unlinkability of the scheme.                                                                                                                                          
\begin{figure}[!h]
    \setlength\abovecaptionskip{0.2\baselineskip}
    \setlength\belowcaptionskip{-2.5pt}
    \centering
    \includegraphics[width=3.1in]{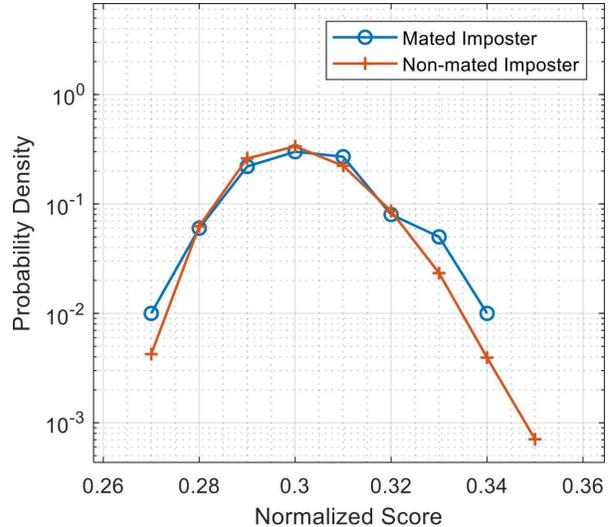}
    \caption{Unlinkability analysis}
\end{figure}\\

\section{Conclusions and Future work}
We have introduced a novel approach for the construction of cancelable and non-invertible fingerprint hash codes using vector permutation and a shift-order process. Particularly, the security mechanism of the scheme against non-invertibility and similarity-based attacks is improved through the integration of shift-order process. Performance accuracy of the scheme outperforms other state-of-the-art recognition systems that comes with, and without the incorporation of security mechanisms. The scheme also satisfy other properties of cancelable biometrics.

Future work will cover security analysis of the scheme against similarity-based attacks, non-invertiblity, and other important security threats. We will also prove that the scheme is able to alleviate the similarity preservation drawback between hash spaces. Consequently, we shall further build upon the shift-ordering procedure by incorporating homomorphic encryption to avert all shortcomings associated with matching in the transformed domain. Instead, matching will be performed in the encrypted domain.

{\small
\bibliographystyle{ieee_fullname}
\bibliography{egbib}

\begin{thebibliography}{10}\itemsep=-1pt

\bibitem{alessandra2013}
Paulino Alessandra~A., Feng Jianjiang, and K.~Jain Anil.
\newblock Latent fingerprint matching using descriptor-based hough transform.
\newblock {\em IEEE Transactions on Information Forensics and Security},
  8(1):31--45, 2013.

\bibitem{andrew2006}
Teoh Beng~Jin Andrew.
\newblock Cancellable biometrics and multispace random projections.
\newblock In {\em 2006 Conference on Computer Vision and Pattern Recognition
  Workshop (CVPRW'06)}, pages 164--164, 2006.

\bibitem{RN24}
Raffaele Cappelli, Matteo Ferrara, and Davide Maltoni.
\newblock Minutia cylinder-code: A new representation and matching technique
  for fingerprint recognition.
\newblock {\em IEEE Transactions on Pattern Analysis and Machine Intelligence},
  32(12):2128--2141, 2010.

\bibitem{charikar2002}
Moses~S. Charikar.
\newblock Similarity estimation techniques from rounding algorithms.
\newblock {\em In Proc. ACM Symp. Theory Computation}, pages 308--388, 2002.

\bibitem{RN25}
Yanzhi Chen, Yan Wo, Renjie Xie, Chudan Wu, and Guoqiang Han.
\newblock Deep secure quantization: On secure biometric hashing against
  similarity-based attacks.
\newblock {\em Signal Processing}, 154:314--323, 2019.

\bibitem{dong2020}
Xingbo Dong, Zhe Jin, Beng Jin~Teoh Andrew, Massimo Tistarelli, and KokSheik
  Wong.
\newblock On the security risk of cancelable biometrics.
\newblock {\em In arXiv preprint}, arXiv:1910.07770v3, 2020.

\bibitem{RN26}
Yi~C. Feng, Meng-Hui Lim, and Pong~C. Yuen.
\newblock Masquerade attack on transform-based binary-template protection based
  on perceptron learning.
\newblock {\em Pattern Recognition}, 47(9):3019--3033, 2014.

\bibitem{RN27}
Matteo Ferrara, Davide Maltoni, and Raffaele Cappelli.
\newblock Noninvertible minutia cylinder-code representation.
\newblock {\em IEEE Transactions on Information Forensics and Security},
  7(6):1727--1737, 2012.

\bibitem{RN28}
Loubna Ghammam, Koray Karabina, Patrick Lacharme, and Kevin Thiry-Atighehchi.
\newblock A cryptanalysis of two cancelable biometric schemes based on
  index-of-max hashing.
\newblock {\em IEEE Transactions on Information Forensics and Security},
  15:2869--2880, 2020.

\bibitem{RN29}
Marta Gomez-Barrero and Javier Galbally.
\newblock Reversing the irreversible: A survey on inverse biometrics.
\newblock {\em Computers \& Security}, 90, 2020.

\bibitem{RN30}
Marta Gomez-Barrero, Javier Galbally, Christian. Rathgeb, and Christopher
  Busch.
\newblock General framework to evaluate unlinkability in biometric template
  protection systems.
\newblock {\em IEEE Transactions on Information Forensics and Security},
  13(6):1406--1420, 2018.

\bibitem{helala2018}
Alshehri Helala, Muhammad Hussain, A.~Aboalsamh Hatim, and A.~Al~Zuair Mansour.
\newblock Cross-sensor fingerprint matching method based on orientation,
  gradient, and gabor-hog descriptors with score level fusion.
\newblock {\em IEEE Access}, 6:28951--28968, 2018.

\bibitem{jain2008}
Anil~K. Jain, Karthik Nandakumar, and Abhishek Nagar.
\newblock Biometric template security.
\newblock {\em Eurasip Journal on Advances in Signal Processing}, 2008.

\bibitem{jain2016}
Anil~K. Jain, Karthik Nandakumar, and Arun Ross.
\newblock 50 years of biometric research: Accomplishments, challenges, and
  opportunities.
\newblock {\em Pattern Recognition Letters}, 79:80--105, 2016.

\bibitem{jin2014}
Zhe Jin, Bok-Min Goi, Andrew Teoh, and Yong~Haur Tay.
\newblock A two-dimensional random projected minutiae vicinity
  decomposition-based cancellable fingerprint template.
\newblock {\em Security and Communication Networks}, 7(11):1691--1701, 2014.

\bibitem{zjin2018}
Zhe Jin, Jung~Yeon Hwang, Yen-Lung Lai, Soohyung Kim, and Andrew Beng~Jin Teoh.
\newblock Ranking-based locality sensitive hashing-enabled cancelable
  biometrics: Index-of-max hashing.
\newblock {\em IEEE Transactions on Information Forensics and Security},
  13(2):393--407, 2018.

\bibitem{RN32}
Zhe Jin, Meng-Hui Lim, Andrew Beng~Jin Teoh, and Bok-Min Goi.
\newblock A non-invertible randomized graph-based hamming embedding for
  generating cancelable fingerprint template.
\newblock {\em Pattern Recognition Letters}, 42:137--147, 2014.

\bibitem{RN33}
Zhe Jin, Meng-Hui Lim, Andrew Beng~Jin Teoh, Bok-Min Goi, and Yong~Haur Tay.
\newblock Generating fixed-length representation from minutiae using kernel
  methods for fingerprint authentication.
\newblock {\em IEEE Transactions on Systems Man Cybernetics-Systems},
  46(10):1415--1428, 2016.

\bibitem{RN34}
Jun~Beom Kho, Jaihie Kim, Ig-Jae Kim, and Andrew Beng~Jin Teoh.
\newblock Cancelable fingerprint template design with randomized non-negative
  least squares.
\newblock {\em Pattern Recognition}, 91:245--260, 2019.

\bibitem{kiran2018}
Raja Kiran~B., Raghavendra Ramachandra, and Busch Christoph.
\newblock Towards protected and cancelable multi-spectral face templates using
  feature fusion and kernalized hashing.
\newblock In {\em 21st International Conference on Information Fusion
  (FUSION)}, pages 2098--2106, 2018.

\bibitem{RN35}
Chulhan Lee, Jeung-Yoon Choi, Kar-Ann Toh, Sangyoun Lee, and Jaihie Kim.
\newblock Alignment-free cancelable fingerprint templates based on local
  minutiae information.
\newblock {\em IEEE Transactions on Systems Man and Cybernetics Part
  B-Cybernetics}, 37(4):980--992, 2007.

\bibitem{RN36}
Kumar Manisha and Nitin Kumar.
\newblock Cancelable biometrics: a comprehensive survey.
\newblock {\em Artificial Intelligence Review}, 53(5):3403--3446, 2020.

\bibitem{RN45}
Takao Murakami, Ryo Fujita, Tetsushi Ohki, Yosuke Kaga, Masakazu Fujio, and
  Kenta Takahashi.
\newblock Cancelable permutation-based indexing for secure and efficient
  biometric identification.
\newblock {\em IEEE Access}, 7:45563--45582, 2019.

\bibitem{nalini2001}
Ratha Nalini~K., H.~Connell Jonathan, and M.~Bolle Ruud.
\newblock Enhancing security and privacy in biometrics-based authentication
  systems.
\newblock {\em IBM Systems Journal}, 40(3):614--634, 2001.

\bibitem{nandakumar2015}
Karthik Nandakumar and Anil~K. Jain.
\newblock Biometric template protection bridging the performance gap between
  theory and practice.
\newblock {\em IEEE Signal Processing Magazine}, 32(5):88--100, 2015.

\bibitem{narishige2015}
Abe Narishige, Yamada Shigefumi, and Takashi Shinzaki.
\newblock Irreversible fingerprint template using minutiae relation code with
  bloom filter.
\newblock In {\em 2015 IEEE 7th International Conference on Biometrics Theory,
  Applications and Systems (BTAS)}, pages 1--7, 2015.

\bibitem{RN37}
Vishal~M. Patel, Nalini~K. Ratha, and Rama Chellappa.
\newblock Cancelable biometrics a review.
\newblock {\em IEEE Signal Processing Magazine}, 32(5):54--65, 2015.

\bibitem{patrick2013}
Lacharme Patrick, Estelle Cherrier, and Christophe Rosenberger.
\newblock Preimage attack on biohashing.
\newblock In {\em 2013 International Conference on Security and Cryptography
  (SECRYPT)}, pages 1--8, 2013.

\bibitem{RN38}
Nalini~K. Ratha, Sharat Chikkerur, Jonathan~H. Connell, and Ruud~M. Bolle.
\newblock Generating cancelable fingerprint templates.
\newblock {\em IEEE Transactions on Pattern Analysis and Machine Intelligence},
  29(4):561--572, 2007.

\bibitem{sani2020}
Abdullahi Sani~M., Wang Hongxia, and Li Tao.
\newblock Fractal coding-based robust and alignment-free fingerprint image
  hashing.
\newblock {\em IEEE Transactions on Information Forensics and Security},
  15:2587--2601, 2020.

\bibitem{scholkopf1998}
B Scholkopf, A Smola, and KR Muller.
\newblock Nonlinear component analysis as a kernel eigenvalue problem.
\newblock {\em Neural Computation}, 10(5):1299--1319, 1998.

\bibitem{RN39}
Muhammad Shahzad, Song Wang, Guang Deng, and Wencheng Yang.
\newblock Alignment-free cancelable fingerprint templates with dual protection.
\newblock {\em Pattern Recognition}, 111, 2021.

\bibitem{RN40}
Amit~Kumar Trivedi, Dalton~Meitei Thounaojam, and Shyamosree Pal.
\newblock Non-invertible cancellable fingerprint template for fingerprint
  biometric.
\newblock {\em Computers \& Security}, 90, 2020.

\bibitem{wang2021}
Hanrui Wang, Xingbo Dong, Zhe Jin, Andrew Beng~Jin Teoh, and Massimo
  Tistarelli.
\newblock Interpretable security analysis of cancellable biometrics using
  constrained-optimized similarity-based attack.
\newblock In {\em Proceedings of the IEEE/CVF Winter Conference on Applications
  of Computer Vision (WACV) Workshops}, pages 70--77, January 2021.

\bibitem{RN41}
Song Wang, Guang Deng, and Jiankun Hu.
\newblock A partial hadamard transform approach to the design of cancelable
  fingerprint templates containing binary biometric representations.
\newblock {\em Pattern Recognition}, 61:447--458, 2017.

\bibitem{RN42}
Song Wang, Wencheng Yang, and Jiankun Hu.
\newblock Design of alignment-free cancelable fingerprint templates with zoned
  minutia pairs.
\newblock {\em Pattern Recognition}, 66:295--301, 2017.

\end{thebibliography}
}

\end{document}